# What Does the Nonextensive Parameter Stand for in Self-Gravitating Systems?

Du Jiulin

*Department of Physics, School of Science, Tianjin University, Tianjin 300072, China*

E-mail: jiulindu@yahoo.com.cn

**Abstract**   It is natural important question for us to ask what the nonextensive parameter stands for when Tsallis statistics is applied to the self-gravitating systems. In this paper, some properties of the nonextensive parameter and Tsallis' equilibrium distribution for the self-gravitating system are discussed in the framework of nonextensive kinetic theory. On the basis of the solid mathematical foundation, the nonextensive parameter can be expressed by a formula with temperature gradient and the gravitational potential and it can be related to the non-isothermal (nonequilibrium stationary state) nature of the systems with long-range interactions. We come to the conclusion that Tsallis' equilibrium distribution is corresponding to the physical state of self-gravitating system at the hydrostatic equilibrium.

# 1. Introduction

The self-gravitating systems have very remarkable physical properties such as inhomogeneity and nonequilibrium due to the long-range nature of the gravitational force. They are thought to be nonextensive. They usually can be at the hydrostatic equilibrium but often cannot be homogeneous and cannot be at thermal equilibrium. In fact, almost all the self-gravitating systems are always non-isothermal if the convective instability is not taking place. The isothermal configurations only correspond to the meta-stable states (locally convective mixing), not true equilibrium states. Therefore, the self-gravitating systems are generally at nonequilibrium states, and Maxwell-Boltzmann (M-B) distribution of the standard statistics can only provide a local description of the properties of the systems because it ignores the spatial inhomogeneity of temperature.

In recent years, it has been considered that the systems with long-range interactions are nonextensive and the conventional Boltzmann-Gibbs (B-G) statistical mechanics may need to be generalized for the statistical description of the features of the systems. The nonextensive generalization of B-G statistical mechanics was made firstly by constructing a new form of entropy with a nonextensive parameter $q$ different from unity (Tsallis, 1988). The resulting "Tsallis statistics" or "Nonextensive Statistical Mechanics (NSM)" has attracted great attention (Abe and Okamoto, 2001). This new theory has been applied to a great number of astrophysical systems and has provided a convenient frame for the thermo-statistical analyses of self-gravitating systems (Tsallis and Prato, 2002), such as stellar polytropes (Plastino and Plastino, 1993; Silva and Alcaniz, 2004), the non-Maxwell velocity distribution of galaxy clusters (Lavagno *et al*, 1998), the gravothermal catastrophe (Taruy and Sakagami, 2002), negative specific heats (Abe, 1999; Silva and Alcaniz, 2003), Jeans criterion for the gravitational instability (Lima *et al*, 2002; Du, 2004a; Du, 2004b), the solar neutrinos (Kaniadakis *et al*, 1996; Lavagno *et al*, 2002; Coraddu *et al*, 2003), and the stellar equilibrium (Du, 2006) and dynamical evolution (Taruya and Sakagami, 2003) etc.

Simultaneously, the understanding of the physical meaning of $q$ has been crucially important in NSM and its applications to the fields of astrophysics. We hope to know the true physical nature of the nonextensive parameter $q$, we hope to know under what circumstances, e.g. which class of nonextensive systems and under what physics situation, should NSM be used for the statistical description, we hope to know what the nonextensive parameter $q\neq 1$ stands for in the self-gravitating systems, and we hope to know why the system with the long-range interparticle force is nonextensive. The very first application of the Tsallis' entropy to astrophysics was done in connection with the stellar polytrope, with the polytropic index n as a function of the nonextensive parameter $q$ in the form of n = 3/2 + 1/($q$-1) (Plastino and Plastino, 1993). Recently, such applications related $q$ to the polytrope index n as n =1/2+ 1/(1-$q$) depending on the new normalized $q$-averages (Taruya and Sakagami, 2002). More generally, Almeida (2001) proved that the canonical distribution function of a thermodynamic system is Tsallis distribution with the distribution function of form $f(\varepsilon) = C[1-(1-q)\beta\varepsilon]^{1/(1-q)}$ if and only if the nonextensive parameter $q$ is expressed by $1-q = d(1/\beta)/d\varepsilon$ with $\beta \propto 1/kT$, where $T$ is the equilibrium temperature and $\varepsilon$ is the energy of the "heat bath". Accordingly, $q \neq 1$ was given a physical interpretation about finite heat capacity of the "heat bath". In addition, Beck (2001) related $q$-1 to the number of degrees of freedom contributing to a fluctuating spatio-temporal temperature field, and Boghosian *et al* (2003) related it to the number of dimensions in lattice Boltzmann models.

In this paper, we will deal with some properties of the nonextensive parameter and the Tsallis distribution for the self-gravitating systems. We discuss the mathematical express of the nonextensive parameter based on the generalized Boltzmann equation in Sec.2. In Sec.3, we analyze the physical meaning of the nonextensive parameter and try to answer the question of what is the physical state when the self-gravitating system is at the Tsallis' equilibrium. The result is naturally obtained by the hydrostatic equilibrium in NSM. In the final section, we give the conclusive remarks.

## 2. Tsallis' Equilibrium Distribution for Self-gravitating Systems

(I) The Generalized Boltzmann Equation

In the framework of NSM, the $q$-generalized kinetic theory was developed recently (Silva, *et al*, 1998; Lima *et al*, 2001). The main result of these works consists in showing that this extended formulation leads to a generalized Maxwellian $q$-velocity distribution. We now consider a system of $N$ particles in the $q$-generalized kinetic theory and the mass of each particle is $m$. The particles can be galaxies, stars, and atoms. We let $f_q(\mathbf{r},\mathbf{v},t)$ be the distribution function of particles in the self-gravitating system, and then $f_q(\mathbf{r},\mathbf{v},t)d^3\mathbf{r}d^3\mathbf{v}$ is the particle numbers at time $t$ and in the volume element $d^3\mathbf{r}d^3\mathbf{v}$ around the position $\mathbf{r}$ and the velocity $\mathbf{v}$. The generalized Boltzmann equation (Lima *et al*, 2001) can be written as

$$\frac{\partial f_q}{\partial t} + \mathbf{v}\cdot\frac{\partial f_q}{\partial \mathbf{r}} + \mathbf{F}\cdot\frac{\partial f_q}{\partial \mathbf{v}} = C_q(f_q) \tag{1}$$

where $C_q$, on the right hande side of this equation, is called the $q$-collision term and is defined by

$$C_q(f_q) = \frac{d^2}{2}\int |\mathbf{V}\cdot\mathbf{e}| R_q(f_q, f'_q) d\omega \, d^3\mathbf{v}_1 \tag{2}$$

with $R_q$ the difference of two correlation functions before and after collision, satisfying the $q$-generalized molecular chaos hypothesis. We now apply the generalized Boltzmann equation to the astrophysical self-gravitating systems. We assume particles to interact via Newtonian gravitation $\mathbf{F} = -\nabla\varphi$, where the gravitational potential $\varphi$ is determined by Poisson equation,

$$\nabla^2\varphi = 4\pi \, Gmn \tag{3}$$

It has been verified that solutions of the generalized Boltzmann equation satisfy the generalized $H$ theorem if $q > 0$ and evolve irreversibly towards the Tsallis' equilibrium distribution,

$$f_Q(\mathbf{r},\mathbf{v}) = nB_Q \left(\frac{m}{2\pi kT}\right)^{\frac{3}{2}} \left[1 - Q\frac{m(\mathbf{v}-\mathbf{v}_0)^2}{2kT}\right]^{\frac{1}{Q}} \tag{4}$$

where $Q = 1 - q$ and the normalization constant is

$$B_Q = \frac{1}{4}Q^{\frac{1}{2}}(2+Q)(2+3Q)\Gamma\left(\frac{1}{2}+\frac{1}{Q}\right)\bigg/\Gamma\left(\frac{1}{Q}\right); \quad \text{for } Q \geq 0.$$

$$B_Q = (-Q)^{\frac{3}{2}}\Gamma\left(-\frac{1}{Q}\right)\bigg/\Gamma\left(-\frac{3}{2}-\frac{1}{Q}\right); \quad \text{for } Q \leq 0.$$

$\mathbf{v}_0$ is the entirety velocity of the system, $n$ is the number density of particles defined by $n(\mathbf{r}) = \int f_Q d^3\mathbf{v}$, and temperature $T$ is considered as a function of $r$. $Q$ is also considered as a function of the position $r$.

(II) The Properties of the Nonextensive Parameter

Some properties of the nonextensive parameter can be analyzed based on the mathematical theory about the generalized Boltzmann equation recently (Du, 2004c). According to the generalized $H$ theorem mentioned above, we have $C_q = 0$ and $\partial f_q / \partial t = 0$ when the system is at the Tsallis' equilibrium. Then the generalized Boltzmann equation (1) is reduced to

$$\mathbf{v} \cdot \nabla f_Q - \nabla \varphi \cdot \nabla_v f_Q = 0 \tag{5}$$

where we have used $\nabla = \partial/\partial \mathbf{r}, \nabla_v = \partial/\partial \mathbf{v}$. Substituting Eq.(4) into Eq.(5), we have

$$\mathbf{v} \cdot \left\{\nabla \ln\left[nB_Q\left(\frac{m}{2\pi kT}\right)^{\frac{3}{2}}\right] + \sum_{i=1}^{\infty}\left(Q^{i-1}\frac{\nabla T}{T} + \left(\frac{1}{i}-1\right)Q^{i-2}\nabla Q\right)\left(\frac{m\mathbf{v}^2}{2kT}\right)^i\right\}$$

$$+ \nabla\varphi \cdot \frac{m\mathbf{v}}{kT}\sum_{i=0}^{\infty}\left(\frac{Qm\mathbf{v}^2}{2kT}\right)^i = 0 \tag{6}$$

Usually, the self-gravitating system at a stable state is believed to satisfy the condition of hydrostatic equilibrium and it is considered as being at a relatively static state. In this case, the macroscopic entirety-moving velocity is zero, $\mathbf{v}_0 = 0$. Furthermore, the coefficients of the powers of **v** in Eq.(6) must be zero because **r** and **v** are independent variables and Eq.(6) is identically null for any arbitrary **v**. The properties of Tsallis' equilibrium distribution for the self-gravitating system can be determined by these coefficient equations (Du, 2004d). One of the properties is related to the nonextensive parameter $q$ by the equation,

$$k\nabla T + Qm\nabla \varphi = 0, \tag{7}$$

which can be written under the spherical symmetry as

$$Q = -\frac{k}{m}\frac{dT}{dr}\bigg/\frac{d\varphi}{dr} = -\frac{k}{\mu m_H}\frac{dT}{dr}\bigg/\frac{GM(r)}{r^2}, \tag{8}$$

where $\mu$ is the mean molecular weight, $m_H$ is the mass of the hydrogen atom, and $M(r)$ is the mass interior to a sphere of radius $r$. It is shown clearly in this relation that the nonextensive parameter is $Q \neq 0$ if and only if the temperature gradient is $\nabla T \neq 0$, which gives a clear physics of $q \neq 1$ with regard to the nature of non-isothermal configurations of the self-gravitating system. If the temperature gradient is $\nabla T = 0$, then we have $Q = 0$ and $q=1$, which corresponds to the case of B-G statistics, while if the temperature gradient is $\nabla T \neq 0$, then we have $Q \neq 0$ and $q \neq 1$, which corresponds to the case of Tsallis statistics. Therefore, the nonextensive parameter $q \neq 1$ is shown to be related to spatial inhomogeneity of temperature in the systems with the gravitational long-range interactions.

(III) The Generalized M-B Distribution

The other property of Tsallis' equilibrium distribution is the temperature-dependent density distribution (Du, 2004c) given by

$$n(\mathbf{r}) = n_0\left(\frac{T(\mathbf{r})}{T_0}\right)^{\frac{3}{2}} \exp\left[-\frac{m}{k}\left(\int \frac{\nabla \varphi(\mathbf{r})}{T(\mathbf{r})} \cdot d\mathbf{r} - \frac{\varphi_0}{T_0}\right)\right], \tag{9}$$

where $n_0, T_0, \varphi_0$ denotes the density, the temperature and the gravitational potential at $r = 0$, respectively. Then, we obtain the generalized M-B distribution (Tsallis' equilibrium distribution) for the self-gravitating system,

$$f_Q(\mathbf{r},\mathbf{v}) = n_0 B_Q \left(\frac{m}{2\pi k T_0}\right)^{\frac{3}{2}} \exp\left[-\frac{m}{k}\left(\int \frac{\nabla \varphi(\mathbf{r})}{T(\mathbf{r})} \cdot d\mathbf{r} - \frac{\varphi_0}{T_0}\right)\right] \left[1 - Q\frac{m\mathbf{v}^2}{2kT(\mathbf{r})}\right]^{\frac{1}{Q}}, \quad (10)$$

The important difference between Eq.(10) and the standard M-B distribution is that Eq.(10) contains the parameter $Q$ and so includes the contributions of temperature gradient $\nabla T$. Due to this characteristic, the generalized M-B distribution could describe the nature of non-isothermal configurations (nonequilibrium stationary state) of the self-gravitating systems. It is clear that the standard M-B distribution can be perfectly recovered from Eq.(10) if we take $\nabla T = 0$. Following this understanding, it is very possible to find the experimental evidence for a value of $q$ different from unity and then to test the theory of nonextensive statistical mechanics.

## 3. The Hydrostatic Equilibrium in NSM

It is natural question for us to ask what is the true physical state of the Tsallis' equilibrium for the self-gravitating system when the nonextensive parameter is related to the nature of nonisothermal configurations in the systems with the long-range interactions. The results about Eq.(7) seem to forebode a relation that exists between the Tsallis' equilibrium distribution and some fundamental dynamical property of the self-gravitating system. We are natural to mention the hydrostatic equilibrium. The hydrostatic equilibrium is one of the basic assumptions for self-gravitating systems as well as one of the fundamental properties about stellar structure and stellar evolution. It is believed that the self-gravitating systems satisfy the condition of hydrostatic equilibrium when they are at stable states and the convective instability is not taking place, the general form of which is written as

$$\nabla P = -mn\nabla \varphi(\mathbf{r}) \quad (11)$$

where $P$ is the pressure of self-gravitating gas. In the framework of Boltzmann-Gibbs (B-G) statistical mechanics, the pressure takes the form of the equation of state of idea gas, $P = nkT$, where the temperature $T$ is constant. But in the framework of NSM, the equation of state of the self-gravitating gas has been modified now. We first analyze the mean value of square velocity of particles,

$$<v^2> = \int_0^{v_{max}} v^2 f_q(\mathbf{r},\mathbf{v})\,d^3v \qquad (12)$$

By using Eq.(10), this mean value can be evaluated easily and the equation of state in NSM is given (Du, 2005) by

$$P = \frac{1}{3}mn<v^2> = \frac{2}{7-5q}nkT, \quad 0 < q < \frac{7}{5}. \qquad (13)$$

Substituting this new equation of state into the equation of hydrostatic equilibrium (11), we have

$$\frac{2}{7-5q}k(n\nabla T + T\nabla n) = -mn\nabla\varphi \qquad (14)$$

where the density gradient $\nabla n$ is related to the temperature gradient and the gravitational potential gradient. From Eq.(9) we find the relation,

$$\frac{\nabla n}{n} = \frac{3\nabla T}{2T} - \frac{m\nabla\varphi}{kT} \qquad (15)$$

Combining Eq.(15) with Eq.(14), we find again the relation of $k\nabla T + (1-q)m\nabla\varphi = 0$, and hence we obtain the formula (8) by using the hydrostatic equilibrium equation. This result tells us that the Tsallis' equilibrium distribution for the self-gravitating system can stand for the physical state of the hydrostatic equilibrium.

## 4. Conclusive Remarks

We have obtained an analytic expression of the nonextensive parameter $q$, which relates $q \neq 1$ to temperature gradient and the gravitational potential, and therefore it describes the nature of the systems with long-range interactions. In the view of mathematics, there seems a question that the potential function $\varphi$ in Eq.(7) can be not only the long-range potential but also the short-range one. However in the view of physics, $\varphi$ can only be the long-range potential. The systems with different interparticle potentials will produce different consequences. If the potential $\varphi$ has the nature of short-range interactions, then each particle in the system is free from the boundary of the system. Such a system is physical "large". The physical "large" system is always to limit to the thermal equilibrium because each particle of the system is free, and temperature gradient becomes zero and $q$ is unity. But if $\varphi$ is the long-range potential, then each particle in the system feels the boundary of the system, and so the system is physical "small". The physical "small" system cannot limit to the thermal equilibrium *by itself* because each particle in the system is not free, and the temperature gradient cannot be to zero and so $q$ is not unity. This is the reason why the stellar interior is always non-isothermal and is always inhomogeneous if the convective mixing has not taken place.

Almost all the systems treated in statistical mechanics with B-G statistics have usually been extensive and at thermal equilibrium, and these properties hold for systems with short-range interparticle forces. The self-gravitating system cannot limit to thermal equilibrium *by itself* but can usually be at the hydrostatic equilibrium due to the long-range nature of gravitational force. In conclusion, B-G statistics can exactly describe the state of the systems with short-range interparticle forces at thermal equilibrium, while Tsallis' equilibrium distribution can describe the state of the systems with long-range interparticle forces at the hydrostatic equilibrium or the nonequilibrium stationary state.

## Acknowledgments

I would like to thank C. Tsallis, S. Abe, P. Quarati, H. J. Haubold and A. Q. Wang for helpful discussions during the twelfth UN/ESA workshop on Basic Space Science at Beijing, 2004, and this paper was reported at the workshop. This work is supported by the project of "985" Program of TJU of China.